\documentclass[prl,twocolumn,amsmath,amssymb,superscriptaddress]{revtex4-1}
\pdfoutput=1

\usepackage[pdftex]{graphicx}
\usepackage[pdftex,colorlinks=true,citecolor=blue,urlcolor=blue,linkcolor=black]{hyperref} 

\usepackage{graphicx}
\usepackage{dcolumn}
\usepackage{bm}
\usepackage{datetime}
\usepackage{color}
\usepackage{float}
\usepackage{ulem}				

\newcommand{\vare}{\varepsilon}

\newcommand{\br}{\textbf{r}}
\newcommand{\bx}{\textbf{x}}

\newcommand{\at}{a_\mathrm{2D}}
\newcommand{\Texp}{T_\textrm{exp}}

\newcommand{\lnkfa}{\ln(k_\mathrm{F}a_\mathrm{2D})}
\newcommand{\EB}{E_\text{B}}

\newcommand{\gr}{g_\text{1}(r)}

\begin{document}

\title{Quantum scale anomaly and spatial coherence in a 2D Fermi superfluid}

\author{P. A. Murthy } 
\email{murthy@physi.uni-heidelberg.de}
\affiliation{Physics Institute, Heidelberg University, Germany}

\author{N. Defenu}
\email{nicolo.defenu@thphys.uni-heidelberg.de}
\affiliation{Institute for Theoretical Physics, Heidelberg University, Germany}

\author{L. Bayha}
\affiliation{Physics Institute, Heidelberg University, Germany}

\author{M. Holten}
\affiliation{Physics Institute, Heidelberg University, Germany}

\author{P. M. Preiss}
\affiliation{Physics Institute, Heidelberg University, Germany}

\author{T. Enss}
\affiliation{Institute for Theoretical Physics, Heidelberg University, Germany}

\author{S. Jochim}
\affiliation{Physics Institute, Heidelberg University, Germany}

\date{\today}

\begin{abstract}
Quantum anomalies are violations of classical scaling symmetries caused by quantum fluctuations. Although they appear prominently in quantum field theory to regularize divergent physical quantities, their influence on experimental observables is difficult to discern. Here, we discovered a striking manifestation of a quantum anomaly in the momentum-space dynamics of a 2D Fermi superfluid of ultracold atoms. We measured the position and pair momentum distribution of the superfluid during a breathing mode cycle for different interaction strengths across the BEC-BCS crossover. Whereas the system exhibits self-similar evolution in the weakly interacting BEC and BCS limits, we found a violation in the strongly interacting regime. The signature of scale-invariance breaking is enhanced in the first-order coherence function. In particular, the power-law exponents that characterize long-range phase correlations in the system are modified due to this effect, indicating that the quantum anomaly has a significant influence on the critical properties of 2D superfluids. 	
\end{abstract}

\pacs{Valid PACS appear here}
\keywords{murthy@physi.uni-heidelberg.de}
\maketitle
Symmetries and their violations are fundamental concepts in physics. A prominent type is conformal symmetry which gives rise to the peculiar effect of scale-invariance, where the properties of a system are unchanged under a transformation of scale. For instance, a Hamiltonian $H(\bx)$ is said to be scale-invariant when $H(\lambda\bx) = \lambda^\alpha H(\bx)$, where $\lambda$ is a scaling factor and $\alpha$ is a real number. Intriguingly, scaling symmetries such as these can be violated by quantum fluctuations, which is known as a quantum anomaly. Such anomalous symmetry breaking is widely discussed in quantum field theory \cite{Weinberg1995}, as they have fundamental implications in a wide range of scenarios, such as high-energy physics and phase transitions. However, experimental signatures of this effect, particularly in many-body systems, have so far been elusive. Here, we report the direct observation of a quantum anomaly in the dynamics of a two-dimensional Fermi superfluid.

Two-dimensional systems with contact interactions, $V(\bx) \propto \delta^2(\bx)$, are particularly interesting in the context of scale-invariance violation, because the $\delta^2$ potential does not introduce a characteristic scale to the Hamiltonian. At the classical level, the transformation $\bx \rightarrow \lambda \bx$  rescales the interaction potential as $V(\lambda\bx)=\lambda^{-2}V(\bx)$ exactly the same way as the kinetic energy and therefore the classical 2D gas is intrinsically scale-invariant \cite{Holstein1993,Pitaevskii1997}. However at the quantum mechanical level, this is no longer true since the $\delta^2$ scattering potential supports a two-body bound state for arbitrarily weak attraction. This additional binding energy scale $\EB$ and the associated scattering length scale $\at$ effectively break the scaling relation between interaction and kinetic energy, which leads to a quantum anomaly.

An important question is, how does this quantum anomaly influence the behavior of 2D systems at macroscopic scales? This is especially relevant for 2D superfluids which exhibit algebraic - hence scale-free - decay of phase correlations \cite{Murthy2015,Hadzibabic2006} described by the Berezinskii--Kosterlitz--Thouless (BKT) mechanism. In this case, how does the introduction of a short-distance scale ($\at$) affect the long-range behavior such as spatial coherence and transport properties in 2D superfluids? These questions are at the heart of many-body physics of 2D systems and answering them may provide insights into the general phenomenology of lower dimensional systems such as exciton-polariton condensates and graphene \cite{Cao2018}.
 
\begin{figure*}[ht!]	
	\includegraphics[scale=1.32]{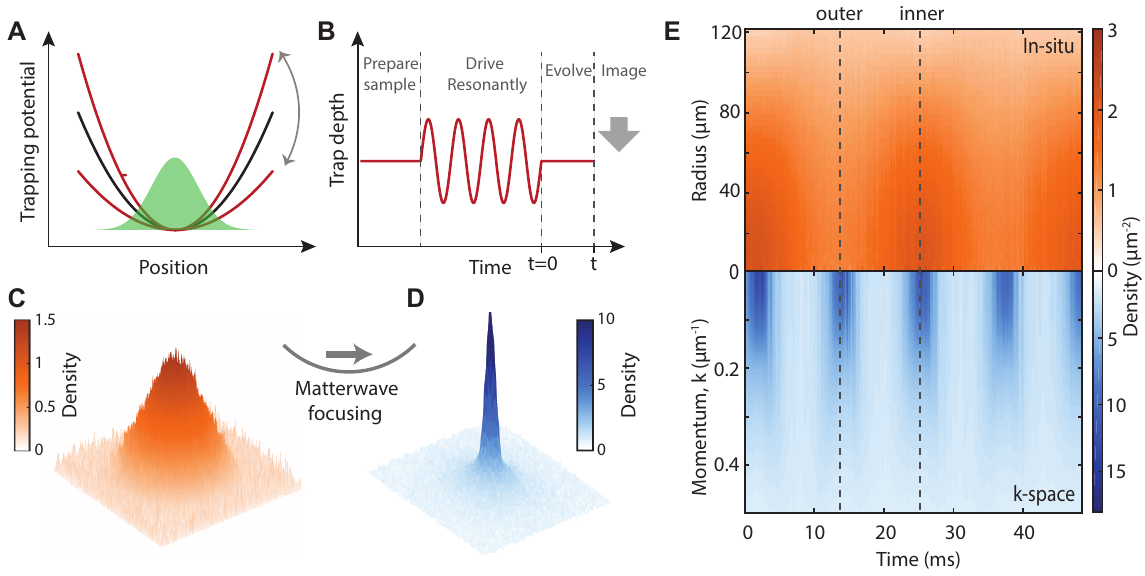}
	\caption{\textbf{Dynamics of a 2D Fermionic superfluid in position and momentum space.} \textbf{A, B} We prepare a 2D Fermi gas well below the superfluid critical temperature \cite{Ries2015}. The isotropic breathing mode is excited by resonantly modulating the harmonic trap. Once the drive is stopped, the breathing oscillations continue for a variable time $t$, at which point we measure (\textbf{C}) the in-situ density distribution $\rho(r,t)$, and (\textbf{D}) the pair momentum distribution $n(k,t)$ using a matterwave focusing technique. \textbf{E} Example of azimuthally averaged $\rho(r,t)$ (orange) and $n(k,t)$ (blue) taken at interaction strength $\lnkfa \approx 1$. The in-situ density oscillates at twice the trap frequency as expected. The momentum distribution exhibits sharp revivals at twice the rate of the in-situ oscillation. The frequency doubling arises from the sinusoidal oscillation of the hydrodynamic velocity field, which vanishes at the inner and outer turning points of the breathing cycle, denoted by the vertical dashed lines.}
	\label{fig:method}
\end{figure*}

In the field of ultracold atomic gases, the issues of scale invariance and quantum anomalies have been previously discussed in literature since interactions between atoms are contact-like to a good approximation. 2D Bose gases in the weakly interacting limit are demonstrably scale-invariant \cite{Hung2011,Desbuquois2014,Pitaevskii1997}, suggesting that the bound state plays a negligible role in these systems. However, in 2D Fermi gases, particularly in the strongly interacting regime, the effect of the additional length scale $\at$ becomes appreciable, for instance in the thermodynamic equation of state \cite{Makhalov2014,Boettcher2016,Fenech2016,Bertaina2011,Shi2015}. On this basis, various theoretical works have predicted a quantitatively pronounced effect of the scale-invariance violation in this regime \cite{Hofmann2012,Moroz2012,Gao2012,Daza2018}. 

Specifically in harmonically trapped gases, a notable manifestation of this anomaly is an interaction-induced correction to the collective monopole frequency with respect to the non-interacting value\,\cite{Olshanii2010,Hofmann2012,Moroz2012,Gao2012,Taylor2012} of twice the trap frequency. Although previous studies on monopole modes found no evidence of such a correction \cite{Vogt2012}, recent experiments have reported the observation of an anomalous frequency shift at low temperatures \cite{Holten2018,Peppler2018}. However, the relative magnitude of these shifts ($\sim 1-2\%$) is several times smaller than the theoretical prediction ($\sim 10\%$), raising questions on the physical relevance of the quantum anomaly for the dynamical properties of 2D Fermi gases.

Here, we discovered that fermionic interactions which lead to the quantum anomaly in fact have a remarkably pronounced influence on the long-range behavior of the 2D system. Rather than the breathing mode frequencies, we explore the spatial coherence properties in momentum space, which reveals the scale-invariance breaking effect that is nearly absent in the position space density profiles. 

In our experiments, we prepared a gas of approximately $2 \times 10^4$ $^6$Li atoms in the lowest two hyperfine states, trapped in a highly anisotropic potential and cooled to low temperatures deep in the superfluid phase. The ratio of absolute temperature to the Fermi temperature $T_F$ is in the range $T/T_F \sim 0.05-0.1$. The radial and axial trap frequencies of the harmonic potential are $\omega_r = 2\pi \times 23\,$Hz and $\omega_z = 2\pi \times 7.1\,$kHz respectively, corresponding to an aspect ratio $\omega_z/\omega_r \approx 310$. With the relevant thermodynamic scales kept smaller than the axial confinement energy, we ensure the system is in the kinematically 2D regime. By tuning the interactions between fermions around a Feshbach resonance, we access the 2D BEC-BCS crossover region. The interactions in the 2D many-body system are described by a dimensionless parameter $\lnkfa$, where $k_\text{F}$ is the Fermi momentum and $a_\text{2D}$ is the 2D scattering length obtained from the 3D scattering length\,\cite{Petrov2001,Dyke2016}. For $\lnkfa \ll -1$, we are in the BEC regime whereas $\lnkfa \gg 1$ corresponds to the BCS regime. The strongly correlated regime located between these limits occurs when $1/k_\text{F} \sim \at$. This crossover region exhibits some intriguing features such as enhanced critical temperature $T_c$ \cite{Ries2015} and a large pseudogap region above $T_c$ where pairing is strongly density-dependent \cite{Murthy2018}.

We investigate the interplay between quantum anomaly and phase correlations by measuring the dynamical evolution of the gas both in position space (i.e. in-situ) and in momentum space. Measuring the momentum distribution is particularly important as it encodes information of phase fluctuations in the superfluid. First, we brought the system out of its equilibrium configuration by resonantly modulating the harmonic trapping potential at twice the trap frequency $2\omega_r$ as illustrated in Fig.\,\ref{fig:method}\,{A, B}. This protocol excites the 2D isotropic breathing mode whereby the gas undergoes periodic cycles of compression and expansion. After a fixed duration (10 cycles), the drive was stopped and the cloud evolved in the original potential for a variable time $t$. In contrast to previous works which investigated the frequency of the breathing mode, we focus on how the shapes of the in-situ and momentum distributions change within a single breathing cycle. Since the damping rate of the breathing mode is very small ($\sim0.01\omega_r$) \cite{Holten2018}, the motion is essentially isentropic which allows to directly probe scale invariant behavior.

To measure $n(k)$, we employed a matterwave focusing technique that has been previously demonstrated for 2D gases \cite{Tung2010,Murthy2014}. First, we rapidly ramped the offset magnetic field to the weakly interacting limit of strongly bound dimers. Immediately following the ramp, we switched off the trapping potential and released the sample to ballistically expand in a shallow harmonic potential for a quarter period $\Texp/4 = \pi/2\omega_\textrm{exp} = 21.8\,$ms, where $\omega_\textrm{exp}$ is the shallow trap frequency. The $\Texp/4$ evolution maps the initial momentum distribution of particles to the spatial distribution. As the time scale of the magnetic field ramp ($\tau_\text{ramp} \sim 50\,\mu$s) is shorter than the intrinsic timescales of the system, the measured spatial distribution at $t=\Texp/4$ reflects to a very good approximation the initial momentum distribution of pairs. The strong enhancement of the low-momentum modes in $n(k)$ as seen in Fig.\,\ref{fig:method}\,{D} signals superfluidity in the system as it is related to long-range spatial coherence in the system \cite{Murthy2015,Ries2015}. 

\begin{figure*}[t!]	
	\includegraphics[scale=1.08]{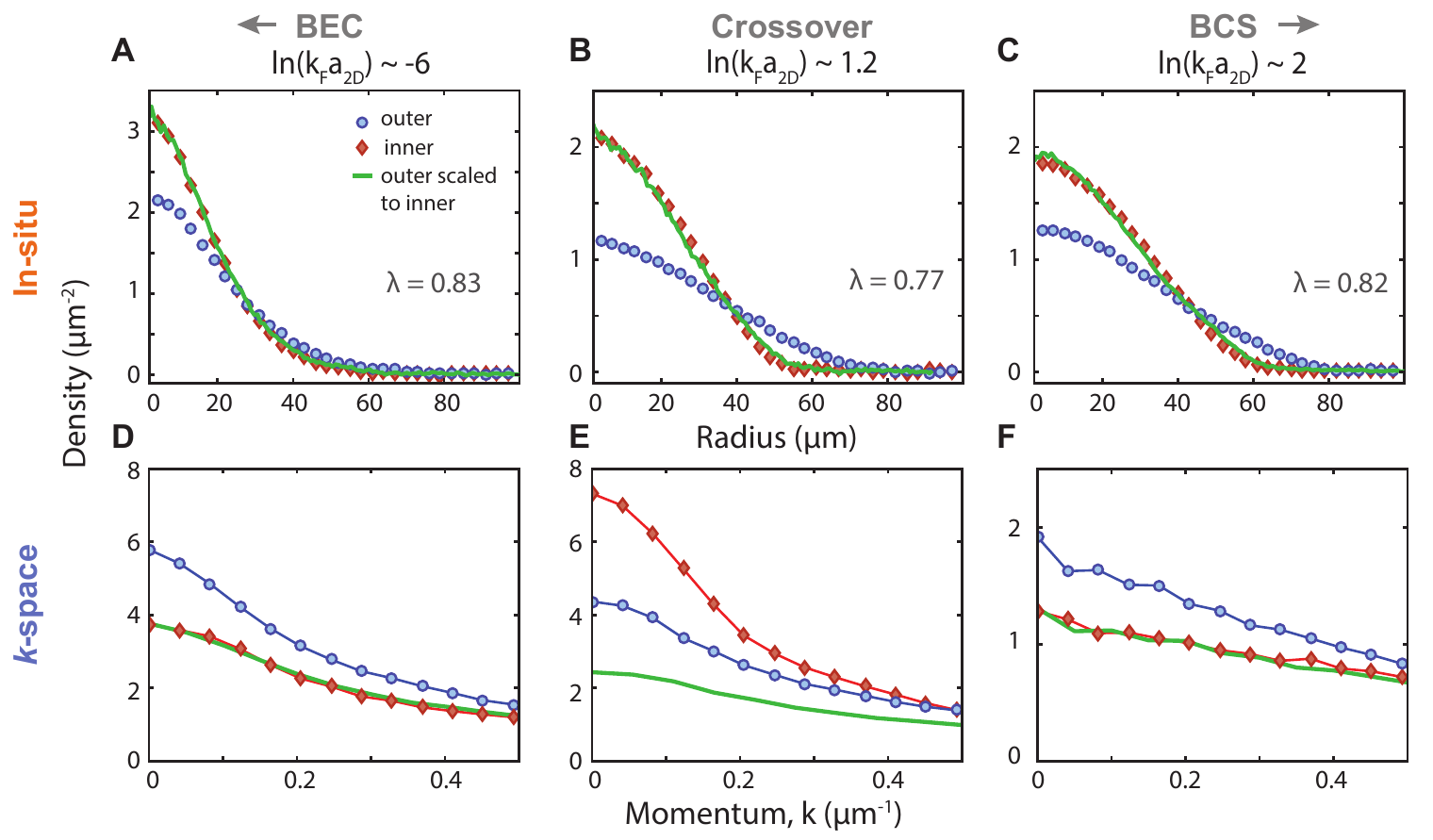}
	\caption{\textbf{Scale-invariance breaking in momentum space.} The in-situ (upper panels) and momentum distributions (lower panels) at the inner and outer turning points for interaction strengths $\lnkfa \approx -6$ (\textbf{A, D}), $1.2$ (\textbf{B, E}) and $2$ (\textbf{C, F}). For a scale-invariant system, the in-situ density profiles at $t_o$ (red diamonds) and  $t_i$ (blue circles) should be scalable with a single scaling factor $\lambda$, as well as the momentum distributions ($n(k,t_o) \rightarrow n(k,t_i)$) with the inverse factor $\lambda^{-1}$. Such scaling behavior is observed both in the weakly interacting BEC and BCS regimes. However in the strongly interacting crossover regime, we find a clear departure from scale-invariance. While the evolution of the $\rho(r)$ is still self-similar (\textbf{B}), the momentum distribution (\textbf{E}) shows a significant discrepancy from the expected result (green). This scaling violation at strong interactions is attributed to the quantum anomaly. Since total particle number is conserved, enhancement of density at low-$k$ is compensated by reduction at high-$k$ (not shown).} 
	\label{fig:compare}
\end{figure*}

In Fig.\,\ref{fig:method}\,E, we show an example of the measured time-evolution of the in-situ $\rho(r,t)$ (orange) and momentum distributions $n(k,t)$ (blue) taken at the interaction parameter $\lnkfa \approx 1$. The in-situ distribution exhibits periodic compression and expansion at approximately twice the trap frequency ($\omega_B \approx 2\omega_r$), as expected. In contrast, $n(k,t)$ undergoes sharp revivals at twice the rate of $\rho(r)$, i.e. when the cloud size is maximum (outer turning point, $t=t_o$) as well as minimum (inner turning point, $t=t_i$). At intermediate time scales between the turning points, $n(k)$ is broadened. At a qualitative level, this peculiar effect can be understood to occur due to the oscillation of the hydrodynamic velocity field, $\mathbf{v_B} \propto \sin(\omega_B t)[x\mathbf{\hat{e}_x} + y\mathbf{\hat{e}_y}] $. During the breathing cycle, $\mathbf{v_B}$ vanishes at the two turning points. At the intermediate points, the non-zero value of $\mathbf{v_B}$ manifests in a broadened momentum distribution with no visible effects in the in-situ profile. We provide a more detailed description of the effect using variational Gross-Pitaevskii computations in \cite{SOM}. A similar effect has been recently predicted for the 1D Bose gas in Tonks-Girardeau regime using scale invariant dynamics\,\cite{Atas2017} and also experimentally observed in the weakly interacting regime\,\cite{Fang2014}. 

From these dynamical measurements, the occurrence and violation of scale invariance can be studied by comparing the in-situ and momentum-space distributions at different points in time. To illustrate this point, let us consider the time-evolution of a scale invariant gas in a harmonic potential. 
Naturally, the presence of a trapping potential naturally introduces a length scale and thus explicitly breaks scale invariance. However, as pointed out by Pitaevskii and Rosch in Ref.\,\cite{Pitaevskii1997}, the special case of a 2D harmonic potential possesses an inherent $SO(2,1)$ symmetry which restores scaling behavior. Consequently, the harmonically trapped scale-invariant gas displays quasi-integrable dynamics with the time-dependent many-body wavefunction being given in terms of the equilibrium one according to
\begin{equation}\label{eq:1}
\psi(X,t) = \frac{1}{\lambda^N}\psi(X/\lambda,t=0)\exp\Bigl(i\frac{m\dot{\lambda}}{2\hbar\lambda}X^2\Bigr)\exp(i\theta(t)),  
\end{equation}
where $X = (\vec{x}_1, \vec{x}_2,\cdots,\vec{x}_N)$ are the $2N$ position coordinates of many-body system, $m$ is the particle mass, $\theta(t)$ is an overall phase, and $\lambda(t)$ is the time-dependent scale factor which obeys the Ermakov-Milne equation \cite{SOM}. From the full wave-function Eq.\,\eqref{eq:1}, one obtains the evolution of the in-situ density and the momentum distribution,
\begin{align}
	\label{eq:2a}	
	\rho(r,t) &= \frac{1}{\lambda^2}\rho\left(\frac{r}{\lambda}, t=0\right); \\ \label{eq:2b}
	n(k,t) &= \lambda^2 \int W\left(\lambda k + 2m\frac{\dot{\lambda}}{\lambda}r,r, t=0\right)d^2r,
\end{align}
in terms of the Wigner function $W(k,r,t)$. Clearly, the in-situ density is completely self-similar (Eq.\,\ref{eq:2a}), i.e. the density at any time $t$ can be rescaled to its initial form using a single scaling factor $\lambda(t)$. When $\dot{\lambda} = 0$, the momentum distribution $n(k,t)$ also displays self-similar scaling with the inverse factor $\lambda^{-1}$. For the breathing modes, $\dot{\lambda} = 0$ at the two turning points. Therefore, a comparison of the in-situ and momentum distributions at the inner and outer turning points can be used as a proxy to study scale invariance.
\begin{figure*}[t!]	
	\includegraphics[scale=1.24]{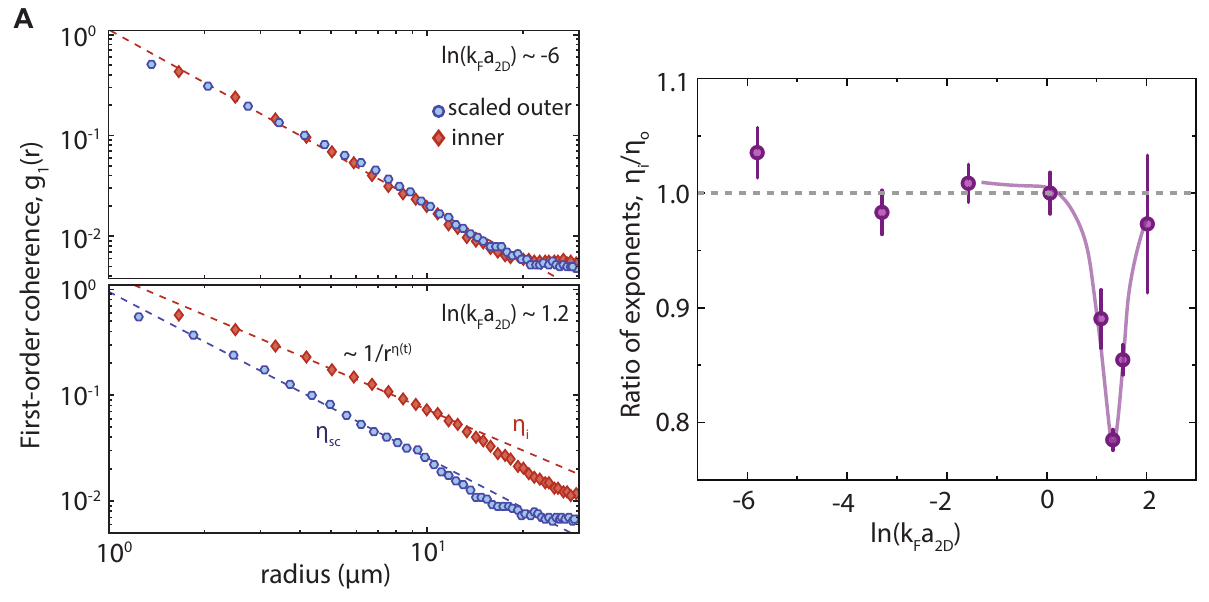}
	\caption{\textbf{The quantum anomaly and spatial coherence.} \textbf{A} The first-order correlation function $g_1(r,t_i)$ at inner point (red) and rescaled correlation function $g_1(\lambda r,t_o)$ at the outer points (blue), for $\lnkfa \sim -6$ (upper panel, BEC) and $\lnkfa \sim 1.2$ (lower panel, crossover). In BEC regime, $g_1(r,t_i)$ and $g_1(\lambda r,t_o)$ coincide, whereas in the crossover regime, the two curves are conspicuously different. From the power-law decay of $\gr \sim r^{-\eta}$, we extract the exponent $\eta$. \textbf{B} The ratio $\eta_i/\eta_o$ across the BEC-BCS crossover. The purple line is a guide to the eye. The scale-invariant expectation $\eta_i/\eta_o = 1$ is reproduced in the BEC regime. In the crossover regime, we observe a sharp dip in the ratio signaling the scaling violation in the long-range phase correlations. The minimum ratio is at $\lnkfa \sim 1.2$ which coincides with the regime of many-body pairing observed in \cite{Murthy2018}.  }
	\label{fig:anomaly}
\end{figure*}

We measured the dynamically evolving in-situ and momentum distributions for various interaction parameters across the BEC-BCS crossover. In Fig.\,\ref{fig:compare}, we show $\rho(r)$ (A, B, C) and $n(k)$ (D, E, F) at the inner and outer turning points for three interaction strengths $\lnkfa = -6,\,1.2,\, \text{and}\, 2$. In the in-situ distributions, we can collapse the $\rho(r,t_o)$ (blue) onto $\rho(r,t_i)$ using a global scaling factor $0< \lambda < 1$. The rescaling is represented by the green curves in panels {A-C}, where $\rho_\text{sc}(r) = \lambda^{-2}\rho(r/\lambda,t_{o})$. The measured and rescaled distributions coincide within the systematic and statistical uncertainties of the measured density, which is approximately $5\,\%$ \cite{Ries2015}.  

In momentum space, the inverse scaling factor $\lambda^{-1}$ should collapse the inner and outer turning point distributions if the system were scaling invariant. This condition is satisfied to a good approximation both in the BEC ($\lnkfa \sim -6$) and BCS ($\lnkfa \sim 2$) regimes (Fig.\,\ref{fig:compare} {D, F} ). In these regimes, the difference between the scaling factor obtained independently for the $k$-space distributions and the inverse in-situ scaling factor is below $2\,\%$. However, in the crossover region $\lnkfa \sim 1.2$,  we find a striking discrepancy between the measured $n(k,t_\text{i})$ at the inner turning point and the rescaled distribution $n_\text{sc}(k)$. In fact, while we expect $n(k,t_i)$ to be always broader than $n(k,t_o)$ (see Fig. \ref{fig:compare} {D,F}), the measured momentum distribution at $\lnkfa \sim 1.2$ shows the opposite effect. Here, the occupation of the low-$k$ region of $n(k)$ is significantly enhanced not only with respect to the expected distribution, but also compared to $n(k,t_o)$. This discrepancy is evidence that scale invariance is violated due to strong interactions, with an unmistakable signature in momentum-space! This is the first main result of this work. 

From Fig.\,\ref{fig:compare}, it is clear that the fermionic interactions have a substantial influence on the low-$k$ modes which correspond to long-wavelength phase fluctuations in the superfluid. The correlations in the phase are characterized by the first-order correlation function 
\begin{equation}
g_1(r) = \int \rho_1(\mathbf{R} - \br/2, \mathbf{R} + \br/2) \, \text{d}\mathbf{R},
\end{equation}
where $\rho_1$ is the one-body density matrix. Experimentally, $g_1(r)$ is directly obtained from the $n(k)$ through a Fourier transform. In our previous work\,\cite{Murthy2015}, we observed the transition from exponential to algebraic decay in $g_1(r)$, in agreement with BKT theory and Quantum Monte Carlo computations \cite{Boettcher2016a}. Here, we use the same procedure described in \cite{Murthy2015} to extract $g_1(r)$ at the inner and outer turning points. These are shown in for Fig.\,\ref{fig:anomaly} {A}, for $\lnkfa = -6$, and $1.2$. To account for the change in cloud size while comparing the two correlation functions, we plot $g_\text{1}(\lambda r,t_o)$ in rescaled coordinates. In addition, we extract the exponent $\eta$ by fitting a power-law ($f(r) \sim r^{-\eta(t)}$) to $g_1(r,t)$. Even though the exponents we measure are larger than the homogeneous BKT predictions, they have the same qualitative behavior \cite{Boettcher2016a}, in particular a smaller exponent corresponds to a larger superfluid phase space density $D_s = \rho_s\lambda_T^2$, where $\rho_s$ is the superfluid density and $\lambda_T$ the thermal de Broglie wavelength.

In the BEC regime, the two curves ($g_1(r,t_i)$ and $g_\text{1}(\lambda r,t_o)$) collapse onto each other (see Fig.\,\ref{fig:anomaly} {A}), whereas in the crossover regime, the correlation functions are substantially different with the inner $g_1(r,t_i)$ decaying slower than expected. In Fig.\,\ref{fig:anomaly} {B}, we show the ratio $\eta_i/\eta_o$ for different interaction strengths across the BEC-BCS crossover. For scale-invariant systems $\eta_i = \eta_o$, i.e the spectrum of phase fluctuations is unaffected by a change in the density. Indeed, we find $\eta_i/\eta_o \approx 1$ in the BEC regime but the ratio dips dramatically in the crossover regime to a value of approximately $0.8$, before rising up again in the weakly interacting BCS regime. This quantitative deviation proves that the quantum scale anomaly that originates in the short-distance fermionic correlations influences the algebraic decay of correlations in the 2D superfluid. This is the second main result of this work. 

What is the origin of these effects? First, we remark that the interaction region ($\lnkfa \sim 1$) where we see the largest scaling violation in the phase correlations coincides with the regions of a) maximum critical temperature \cite{Ries2015}, b) largest density-dependent pairing (pseudogap) \cite{Murthy2018} and c) the maximum breathing mode frequency shift \cite{Holten2018}. This suggests that all these effects may have a common mechanism. One possible explanation comes from the density-dependent pairing effect observed in \cite{Murthy2018}. Specifically in the crossover region, a change in density during the breathing cycle corresponds to a change in the total pairing energy in the system. 
The spatial coherence is carried by fermion pairs of fixed size $\at$, so at the inner turning point where the particle spacing is smallest, more of these pairs overlap. This implies enhanced phase coherence extending over more particle spacings, and a smaller decay exponent $\eta$. At the same time, enhanced occupation of low-momentum modes requires, at fixed total number, a reduced occupation at high momenta and hence a depletion in the pair kinetic energy. We have analyzed the kinetic energy extracted from the momentum distribution and indeed found a scaling violation consistent with this argument \cite{SOM}.



The observations in Fig.\,\ref{fig:anomaly} may also provide hints towards explaining the enhanced critical temperatures in this region. We recall that the power law exponents are an indicator of superfluid stiffness and phase space density: a smaller $\eta$ corresponds to more coherence and larger $D_s$. For scale invariant systems, $D_s$ necessarily remains constant throughout the breathing cycle leading to $\eta_i/\eta_o = 1$. However in the crossover regime, the observation of $\eta_i/\eta_o < 1$ implies that the density-dependent pair correlations in fact enhance the superfluid phase space density for the same effective temperature. In other words, the critical $D_s$ required for the superfluid transition can be attained at higher $T_c/T_F$, as seen in \cite{Ries2015}.

Finally, we highlight some points that may be relevant for future investigations on this topic. First, the density profile does not exhibit significant effects of scale-invariance and satisfies the prediction of the dynamical $SO(2,1)$ symmetry \cite{Taylor2012}. This is consistent with the small shifts in the breathing mode frequency recently reported in \cite{Holten2018,Peppler2018}. It also shows that the breathing dynamics are not fully explained by the measured equation of state \cite{Makhalov2014,Boettcher2016,Fenech2016}, which is scale dependent and would imply a large shift in the breathing mode frequency accompanied by an observable change in the in-situ density profile. 



In momentum space, we found that short-distance fermionic correlations which break scale-invariance have a significant impact on the low-momentum modes, which correspond to the long-wavelength phase fluctuations in the superfluid \cite{Ries2015,Murthy2015,Dalibard2011}. This implies that phenomena fundamentally connected to the phase fluctuations, such as transport, are influenced by the quantum anomaly. 


\newpage
\noindent
\textbf{Acknowledgements} We are grateful to I. Boettcher, T. Gasenzer, J. Hofmann, K. V. Kheruntsyan, T. Lompe and S. Moroz for insightful discussions. This work has been supported by the ERC consolidator grant 725636, the Heidelberg Center for Quantum Dynamics and is part of the DFG Collaborative Research Centre €œSFB 1225 (ISOQUANT). P.M.P. acknowledges funding from European Union's Horizon 2020 programme under the Marie Sklodowska-Curie grant agreement No. 706487. Supporting data can be found in the Supplementary Material. Raw data is available upon request.\\

\noindent
\textbf{Author contributions} P.A.M performed the measurements and data analysis. N.D and T.E provided the theoretical description of the observations. L.B, M.H and P.M.P assisted with experiments and interpretation of data. T.E and S.J supervised the project. P.A.M and N.D conceptualized the manuscript and contributed equally to this work.



%
\providecommand{\noopsort}[1]{}\providecommand{\singleletter}[1]{#1}%
%

%
%
%
%
%
%
%
%
%
%
%
%
%
%
%

\clearpage
\begin{center}
\textbf{\large Supplemental Materials: Quantum  scale anomaly and spatial coherence in a 2D Fermi superfluid}
\end{center}
\setcounter{equation}{0}
\setcounter{figure}{0}
\setcounter{table}{0}
\setcounter{page}{1}
\makeatletter
\renewcommand{\theequation}{S\arabic{equation}}
\renewcommand{\thefigure}{S\arabic{figure}}
\renewcommand{\bibnumfmt}[1]{[S#1]}
\renewcommand{\citenumfont}[1]{S#1}
\section{Materials and Methods}

\subsection{Preparing the sample}
We start our experiments with a molecular Bose--Einstein condensate of approximately 50,000 atoms in the two lowest hyperfine states of $^6$Li, which are prepared after a sequence of optical evaporative cooling at at magnetic offset field of 795 G. Thereafter, we transfer the atoms into an optical standing wave trap (SWT) that is created by interference between two cylindrically shaped far detuned laser beams (1064 nm) at a shallow angle of 14$^{\circ}$. The spacing between the interference fringes is approximately $4\,\mu$m which allows us to load more than 95$\%$ of the atoms into a single layer. In the SWT, we perform additional evaporative cooling which results in a gas of $\sim 2 \times 10^4$ atoms at a temperature of $60\,$nK ($T/T_\text{F} \approx 0.05$). The experimental system and protocol for preparing the sample have been discussed in detail in our previous work \cite{Ries2015}.

\subsection{Frequency doubling in $k$-space}
The pair momentum distribution $n(k)$ displays sharp revivals both at the inner and the outer turning points of the breathing dynamics. Similar behavior was previously predicted in the oscillatory motion of a 1D Bose gas in the Tonks-Girardeau (TG) limit\,\cite{Atas2017}. The latter one dimensional system is integrable and the dynamics could be computed exactly. 

\begin{figure}[t!]
	\includegraphics[]{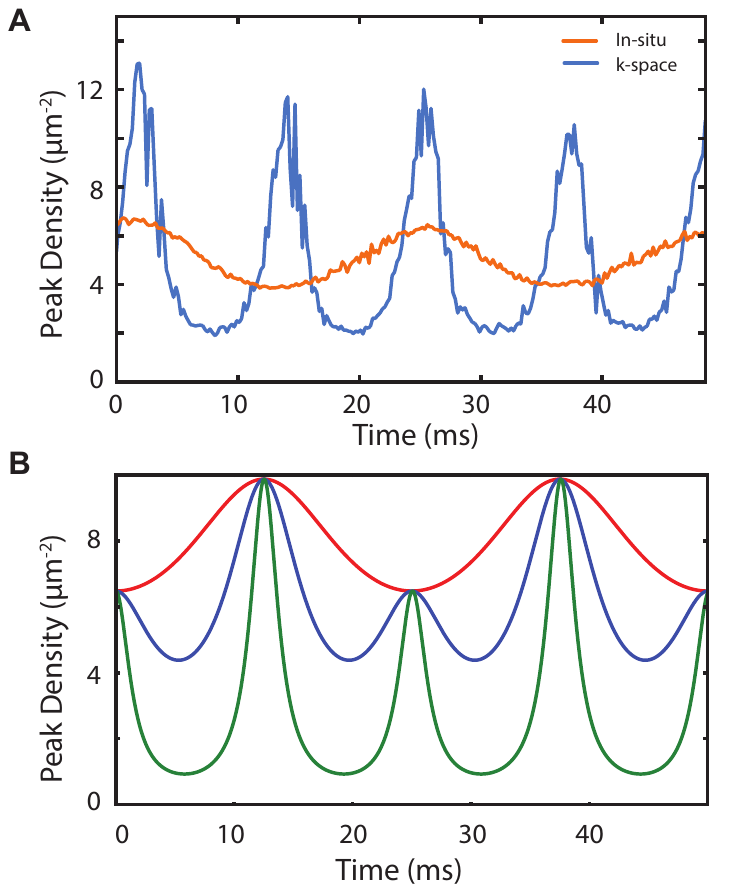}
	\caption{Panel \textbf{A}: Experimentally measured in situ width (orange curve) and zero momentum density (blue curve) as a function of time. Panel \textbf{B}: Variational analysis of the zero momentum spectral density as a function of time for increasing interaction strength $\tilde{g}=0,1,10$ in red, blue, green respectively ($\delta_{0}=-0.01\bar{w}$). Given the present dynamical protocol, frequency doubling appears for large interaction strengths, due to the depletion of low momentum occupancies in the time interval between the breathing turning points, see panel \textbf{B}. The threshold value for the interaction strength above which the frequency doubling appears decreases increasing the quench width $\delta_{0}$.}
	\label{Fig1}
\end{figure}
In the present two dimensional case one shall rely on an approximate procedure. In the BEC limit $\text{ln}(k_{\text{F}}a_{2\text{D}})\ll -1$ the system behaves as a weakly interacting $2D$ Bose-Gas. Ignoring finite temperature effects the system can be described by the celebrated Gross-Pitaevskii equation (GPE)
\begin{align}
\label{GPE}
i\hbar\partial_{t}\psi=-\frac{\hbar^{2}}{2m}\nabla^{2}\psi+V(\boldsymbol{r})\psi+\frac{g}{m}|\psi|^{2}\psi
\end{align}
where $m$ is the Bosonic molecule mass and $g\approx -2\pi/\log(a_{2D}/l_{z})$, where $l_{z}$ is the oscillator length in the transverse direction. This nonlinear equation describes the evolution of the macroscopic coherent wave function $\psi$ at low temperatures.  It is possible to approximately solve the GPE by means of a variational analysis\,\cite{Perez-Garcia1996}. In order to exactly reproduce the non-interacting solution in the $g\to 0$ limit one shall assume a Gaussian trial wave-function
\begin{align}
\label{trial_wf}
\psi(r)=A(t)\prod_{\mu=x,y}&\Biggl{(}e^{-\frac{\left(r_{\mu}-r_{0\mu}(t)\right)^{2}}{2w_{\mu}(t)^{2}}+i\left(r_{\mu}-r_{0\mu}(t)\right)\alpha_{\mu}(t)}\nonumber\\
\cdot
&e^{i\left(r_{\mu}-r_{0\mu}(t)\right)^{2}\beta_{\mu}(t)}\Biggr{)}.
\end{align}
At a given time, this function defines a Gaussian distribution centered at the position $(r_{0x},r_{0y})$, where the transverse $z$ direction has been discarded, since it is tightly confined. The other variational parameters are $A$ (amplitude), $w_{\mu}$ (width), $\alpha_{\mu}$
(slope), and $\beta_{\mu}$, where $\mu\in \{x, y\}$ is an index labeling the spatial dimensions. All the parameters are real numbers  and the amplitude evolution follows by the normalization condition for the wave-function
\begin{align}
A(t)=\sqrt{N}\sqrt{\pi w_{x}(t)w_{y}(t)}^{-1}
\end{align}
where $N$ is the occupation number of the coherent state $\psi$. 
The imaginary terms appearing in the exponent of Eq.\,\eqref{trial_wf} represent the conjugate momenta of the width $w$ and center of mass coordinate $\alpha$. Within this ansatz one obtains the following analytic expressions for the in-situ density
\begin{align}
n(\boldsymbol{r})=|\psi(\boldsymbol{r})|^{2}=\pi^{-1}\prod_{\mu=x,y}\frac{e^{-\frac{\left(r_{\mu}-r_{0\mu}(t)\right)^{2}}{w_{\mu}(t)^{2}}}}{w_{\mu}(t)},
\end{align}
and the spectral distribution
\begin{align}
\label{spec_den}
n(k)=(4\pi)\prod_{\mu=x,y}\frac{w_{\mu}e^{-\frac{w_{\mu}(t)^{2}(\alpha_{\mu}(t)-k)^{2}}{1+4\beta_{\mu}(t)^{2}w_{\mu}(t)^{2}}}}{\sqrt{1+4\beta_{\mu}(t)^{2}w_{\mu}(t)^{4}}}\end{align}

The motion equations of the variational parameters follow from the minimization of the semi-classical action of the system within the restricted trajectory space parametrized by ansatz\,\eqref{trial_wf}\,\cite{Perez-Garcia1996}
\begin{align}
\alpha_{\mu}&=m\,\dot{r}_{0,\mu}\label{eqm2}\\
\ddot{r}_{0\mu}&=-m\omega_{r}^{2}r_{0\mu}\label{eqm1}\\
\beta_{\mu}&=-\frac{m}{2}\frac{\dot{w}_{\mu}}{w_{\mu}}\label{eqm3}\\
\ddot{w}_{\mu}+\omega_{r}^{2}w_{\mu}&=\frac{1}{m^{2}}\frac{1}{w_{\mu}^{3}}+\frac{\tilde{g}}{w_{\mu}\prod_{\alpha}w_{\alpha}}\label{eqm4}
\end{align}
where $\tilde{g}$ is an effective coupling obtained rescaling the microscopic coupling $g$ by a coefficient proportional to the occupation number of the coherent state $\psi$. In the traditional GPE perspective one expects a macroscopic occupation of the ground state wave-function hence $\tilde{g}\gg g$, however in 2D phase fluctuations are divergent at finite temperature and the effective $\tilde{g}$ value can be considerably smaller $\tilde{g}\approx g$\,\cite{Hadzibabic2009}. 

The dynamics analyzed in the paper is equivalent to a sudden quench of the width parameter. Therefore we solve the equations of motion with initial condition $w_{x}=w_{y}=\bar{w}+\delta_{0}$ where $\bar{w}$ is the equilibrium width ($\bar{w}=\omega_{r}^{-1/2}+O(g)$) and $\delta_{0}$ is a finite displacement. In the harmonic approximation $\delta_{0}\ll \bar{w}$ one obtains $w(t)=\bar{w}+\delta_{0}\cos(2\omega_{r}t)$. As a consequence the zero momentum component of the spectral density becomes
\begin{align}
\label{zero_mom_den}
n_{k=0}=\frac{(4\pi)^{3/2}(\bar{w}+\delta_{0}\cos(2\omega_{r}t))^{2}}{1+\delta_{0}^{2}\omega_{r}^{2}\sin(2\omega_{r}t)^{2}(\bar{w}+\delta_{0}\cos(2\omega_{r}t))^{2}}.
\end{align}
where the units were chosen such that $m=1$. The analysis of Eq.\,\eqref{zero_mom_den} explains all the features of the observed frequency doubling: the numerator has only one maximum, for $2\omega_{r}t=0$ and one minimum for $2\omega_{r}t=\pi$, and produces the expected oscillatory behavior due to periodic compression and expansion of the cloud density. 

However, the simple behavior of density oscillations  is
 modified by the phase contributions of the denominator in Eq.\,\eqref{zero_mom_den}, see also Eqs. \eqref{trial_wf} and \eqref{spec_den}. The denominator in Eq.\,\eqref{zero_mom_den} has two minima, for $2\omega_{r}t=0\,\,{and}\,\,\pi$, one in correspondance to the maximum of the numerator the other to the minimum. Therefore the harmonic approximation is consistent with the appearance of \emph{two} maxima of the zero momentum spectral density for each period $T=\pi/\omega_{r}$ of the breathing oscillations.

Since the equilibrium value of the width $\bar{w}$ increases with $g$, the possibility of having two maxima within one single breathing period is regulated by the strength of the interaction, at least in the harmonic approximation. Indeed, in the limit $\delta_{0}/\bar{w}\ll 1$ only the denominator in \eqref{zero_mom_den} contributes to the zero momentum density, as it is shown in Fig.\,\ref{Fig1}.  
In the case $\delta_{0}\approx \bar{w}$ the harmonic approximation  is not valid and we cannot employ formula \eqref{zero_mom_den}. Still a more careful analysis shows that the doubling effect is present and it appears already at smaller interaction strengths.

The comparison between this theoretical picture and the experimental data is reported in Fig.\,\ref{Fig1}. For small quenches $\delta_{0}=0.01\bar{w}$ and large interactions strength $g\approx 30$ (green curve in panel {B}) we find rather good agreement with the measured oscillations of the zero momentum component in the crossover regime (blue curve in panel {A}), consistently with the expectation of fermionic dynamics being described by GPE on the BEC side of the crossover. 

The variational approach depicted above is consistent with the scale invariant dynamics observed in 2D and described in the main text, as long as the position space Gaussian profile \eqref{trial_wf} is replaced with a generic rescaled many body wave-function $\Psi(X/\lambda(t))$, where the time dependent scale parameter $\lambda$ obeys the Ermakov-Milne equation 
\begin{align}
\ddot{\lambda}+\omega_{r}^{2}\lambda =\frac{\omega_{r}^{2}}{\lambda^{3}},\end{align}
with $\lambda=1$ at the beginning of the dynamical evolution.

\begin{figure}[b!]	
	\includegraphics[scale=0.35]{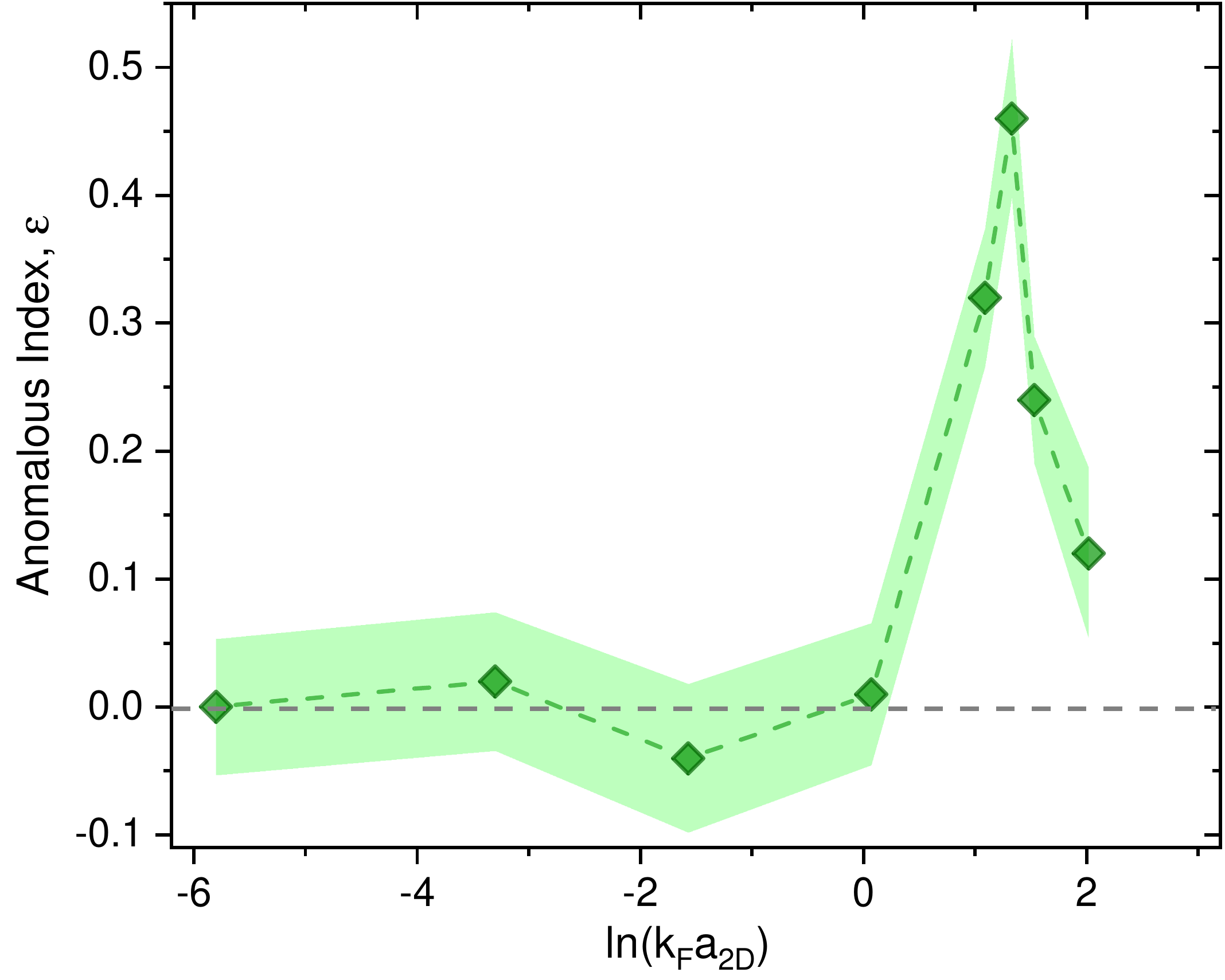}
	\caption{\textbf{Anomalous index.}} 
	\label{fig:anomaly2}
\end{figure}

%

\subsection{Kinetic energy scaling}
Here, we estimate the extent of scale invariance breaking in the system by considering the evolution of the pair kinetic energy ($T$), which is obtained from the instantaneous momentum distributions according to $T(t) = (\hbar^2/2m)\int n(k,t)k^2\text{d}^2k/(2\pi)^{2}$, where $m$ is the dimer mass. Specifically, we define the dimensionless parameter, which we refer to as the anomalous index
\begin{equation}
\label{an_in}
\vare= \frac{\log(T_o/\lambda^2T_i)}{\log(\lambda^2)},
\end{equation}
where $T_i$ and $T_o$ are the kinetic energies at the inner and outer points. The logarithmic derivative allows to quantitatively estimate the scaling violation in the kinetic energy, independent of the absolute energy scale in the system. In the scale invariant case, $T_i = T_o/\lambda^2$ and hence $\vare = 0$, whereas $\vare \neq 0$ in the presence of quantum anomalous corrections which modify the scaling of kinetic energy, i.e. $T_i = T_o/\lambda^2 \rightarrow T_o/\lambda^{2 - 2\vare}$. 

The measured values of $\vare$ across the BEC-BCS crossover are shown in Fig.\,\ref{fig:anomaly}. We find $\vare \approx 0$ in the BEC regime upto $\lnkfa \approx -0.5$. In the crossover regime, the inner kinetic energy is observed to be significantly smaller than expected (i.e $ T_i < T_o/\lambda^2$), and therefore $\vare$ is positive with a peak value of $ \vare \approx 0.4$ at $\lnkfa \sim 1.2$. For weaker interaction strengths in the BCS regime(larger $\lnkfa$), $\vare$ shows a declining trend towards the scale invariant value. Intriguingly, this regime of scaling violation ($\lnkfa \approx 1$) coincides very closely to region where we previously observed the many-body pairing in the system \cite{Murthy2018} as well as the maximum shift in the breathing mode frequency \cite{Holten2018}.

\clearpage

\end{document}